\documentclass[12pt,epsf]{article}
\usepackage{graphicx}
\usepackage{amsmath,amssymb}
\usepackage{cite}

\usepackage{subfigure}

\begin{document}
\begin{titlepage}
\begin{center}

 \vspace{-0.7in}

{\large \bf Finite-Size Effects\\
\vspace{.06in} in \\
\vspace{.06in} Disordered $\lambda\varphi^{4}$ Model}\\
\vspace{.5in}
{\large\em
R. Acosta Diaz\,\footnotemark[1] and N. F. Svaiter\,\footnotemark[2]}\\
\vspace{.06in}
Centro Brasileiro de Pesquisas Fisicas-CBPF
\\ Rua Dr. Xavier Sigaud 150
Rio de Janeiro, RJ,22290-180, Brazil\\

\subsection*{\\Abstract}
\end{center}

We discuss finite-size effects in one disordered $\lambda\varphi^{4}$ model defined in
a $d$-dimensional Euclidean space.
We consider that the scalar field satisfies periodic boundary conditions 
in one dimension and it is coupled with a quenched
random field. 
In order to obtain the average value of the free energy of the system we use the
replica method.
We first discuss
finite-size effects in the one-loop approximation in $d=3$ and $d=4$. We show that in both cases there is a critical length
where the system develop a second-order phase transition, when the system presents long-range correlations with
power-law decay. 
Next, we improve the above result studying the gap equation for the size-dependent squared mass,
using the composite field operator
method. We obtain again, that the system present a second order phase transition with long-range correlation with power-law decay.

\vspace{.06in}

\footnotetext[1]{e-mail:\,\,racosta@cbpf.br}
\footnotetext[2]{e-mail:nfuxsvai@cbpf.br}

PACS numbers:03.70+k,04.62.+v

\end{titlepage}
\newpage\baselineskip .18in

\section{Introduction}
\quad

Disordered systems has been investigated for decades in statistical mechanics, condensed matter
and even gravitational physics \cite{pe2,pe10,pe11,dis1,dis2,dis3}. An example of a system in statistical
mechanics with disorder is the random field Ising model, introduced in the literature by Larkin in the
70's \cite{larkin}.
The hamiltonian of the random field Ising model is analogous to the one of the
classical Ising model, but allowing for a quenched random magnetic field \cite{binder,livro1,livro2,livro3,livro4}.
Some realizations of random field systems are diluted frustrated antiferromagnets \cite{afm} and 
binary liquids in porous media \cite{fluids1,fluids2}. 
The random field Ising model can be used as a model in the case of binary fluids confined in porous media, 
when the pore surfaces couple differently to the two components of a phase-separating mixture. 
%
After  intensive investigations, the properties of the phase transition of the random field Ising model is still under debate.
The first question about the model is its lower critical dimension.
Intending to extend the Peierls argument to systems with symmetry breaking randomness, Imry and Ma \cite{ma} obtained that
the model with nearest neighbour interaction presents spontaneous magnetization only for $d\geq \,3$. This result,
that in $d=2$ the model does not present any spontaneous magnetization is in contradiction with the dimensional
reduction argument, elaborated by Parisi and Sourlas \cite{sourlas,parisi1}.
The controversy was solved by Bricmont and Kupiainen \cite{bric1,bric2} and Aizenman and Wehr \cite{wehr1}. The first authors
proved that there is a phase transition in the random field Ising model for $d\geq\,3$, and the second ones proved that
for $d=2$ there is no phase transition in the model. 
Although these new important results solved the controversy concerning to the
lower critical dimension of the model, phase structure at low and high temperatures, i.e.,
the phase diagram of the model is still under debate. 

Using statistical field theory methods
many papers discussed this model \cite{mezard1, mezard2, dotsenko, orland,dotsenko2, dotsenko3,sherington1,sherington}, based in the fact that
many problems in the theory of phase transitions 
can be formulated in the language of Euclidean field theory. The generating functional of Schwinger functions of a Euclidean field theory
can be interpreted as the partition function of a $d$-dimensional
classical statistical field theory. The zero temperature limit
of this theory corresponds to the usual infinite volume classical partition
function. Finite temperature effects in the generating functional of Schwinger functions
corresponds to a finite size in one direction for the classical partition
function. Our aim here is to use this Euclidean formalism of field theory 
to study finite size effects in classical statistical systems with disorder.

A continuous description instead dealing with discrete elements for the random field 
Ising model is given by the scalar field Landau-Ginzburg model,
where the order parameter is a continuous field $\varphi(x)$, coupled with a quenched random field $\mu(x)$.
In the presence of the disorder, ground state configurations of the
field $\varphi(x)$ are defined by a saddle-point equation, where
the solutions of such equation depends on particular configurations of the quenched random fields.
The existence of these several local minima make very difficult to implement a perturbative approach in a straightforward way.
An alternative route to
circumvented this problem is to average the free energy over the random field, using the replica method \cite{edwards}.
One start
with a system with quenched disorder, where all the correlation
functions are not translational invariant. To calculate the average free energy
a new system defined by $n$ statistically independent replicas is introduced.
After integrating out the random field, it is obtained a
system where the correlation functions are translational invariant.
In this case, these replicas are not more statistically independent and 
the average free energy of the system can be calculated from a zero-component
field theory.

The aim of this paper is to discuss the phase transition in one classical statistical system with disorder, being more specific the $\lambda\varphi^{4}$ model,
considering  that the scalar field satisfies periodic boundary conditions 
in one dimension. Finite size-systems in classical statistical mechanics is any system which has a finite
size in at least one dimension \cite{danchev}.
We would like to point out that the renormalization procedure in a scalar quantum field with annealed random mass was investigated in Ref. \cite{nami}.
There are two interesting cases, the film geometry with $L\times\,\mathbb{R}^{d-1}$ and $S^{1}\times\,\mathbb{R}^{d-1} $.
In this work we assume boundary conditions which do not break the translational symmetry of the replica field theory.
This choice avoid surface effects, which introduce surface divergences in the theory.

Suppose a scalar quantum field theory without disorder, at zero temperature and the system obeys periodic boundary condition in one space dimension.
For very small radius, since there is a size-dependent renormalized mass, the system
present finite correlation length.
This mass generation in the $\lambda\varphi^{4}$ theory has also been noted and studied by Ford and 
Yoshimura \cite{yoshimura} and Birrel and Ford \cite{birrel}.
Using this mechanism in a classical statistical system with disorder, increasing the radius of the compactified dimension, we
obtain the critical size, where a zero-temperature second-order phase transition occurs.
The system presents long-range correlations with
power-law decay.
We would like to stress that
this program to study finite size effects in models with randomness is not new in the literature.
For example, in the Sherington-Kirkpatrick spin glass, finite size effects in the model was investigated in
Refs. \cite{s1,s2}. Also, more recently finite size-effects  was discussed in two papers
\cite{ricci1,ricci2}. In these papers the authors study finite-size corrections to the free energy density in disordered
spin systems on sparse random graphs. Also, finite size effects in quantum field theory,
was discussed for example in the refs. \cite{finitez1,finitez2,finitez3,finitez4}.

In this paper we are studying a scalar field theory with a
$(\lambda\varphi^{4})$ self-interaction defined in a
$d$-dimensional Euclidean space in the presence of disorder.
In this field theory defined in the continuum there is no intrinsic short distance or
a cut-off. An analytic regularization and renormalization program can be easily implemented. As we expected, all the counterterms
that we have to introduce are independent of the radius of the compactified dimension.
Studying the gap equation in the one loop-approximation, it is possible to estimate
the critical size where the system develop a second-order phase transition. Some papers discussing
generation of thermal mass in field theory in the perturbative and non-perturbative regime
are the Refs. \cite{ford,adolfo1,khanna}.
Next, we obtain non-perturbative effects using the composite operator formalism \cite{jackiw,ananos,gap,jpa}.
In this approach one consider a generalization of the effective action, which
depends on the expectation value of the field and the expectation value of the products of two fields \cite{kap,dru,dirk}.
An alternative way to obtain non-perturbative results is to use the Dyson-Schwinger equations. These equations are an infinite
tower of coupled equations. One way to treat this problem is to assume a truncation. In the weak coupling expansion,
it can be shown that the Dyson-Schwinger equations contain perturbation theory. Other way to obtain non-perturbative  results is to use the 
renormalization group equations. See, for example the Refs. \cite{novo1,novo2,novo3}.

The organization of the paper is as follows:
In section II we discuss the composite field operator method for the $\lambda\varphi^{4}$ model.
In section III we study the replica field theory in the disordered scalar field model.
In section IV we study finite-size effects in the $\lambda\varphi^{4}$ model with replicas.  We study this model in a
$d$-dimensional space with one compactified dimension using the
one-loop approximation and also we obtain non-perturbative results using the Dyson-Schwinger equations. Conclusions are given
in section V. To simplify the
calculations we assume the units to be such that
$\hbar=c=k_{B}=1$.

\section{The composite field operator method}
\quad
In the formal functional integral formulation there are two kinds of random  variables. The first one are
the Euclidean fields $\varphi_{i}(x)$. These fields describes generalized Euclidean processes
with zero mean and covariance $S(x-y,m_{0})=\langle\,x|(-\Delta+m_{0}^{2})^{-1}|\,y\rangle$. There are also variables
with the absence of any gradient $\mu(x)$ that makes or the field statistically independent in every point of
the domain or for field that are not statistically independent for different points
of the domain, the two-point correlation function is not defined in terms of gradients.
We call these as uncorrelated and correlated random fields respectively.
We are studying a field theory describing a scalar field with a
$(\lambda\varphi^{4})$ self-interaction defined in a
$d$-dimensional Euclidean space, coupled with a quenched random field. The definition of quenched variables will be clarified latter.
Finite-size effects will be introduced in field theory assuming that one dimension is compactified,
where the radius of the compactified dimension is $L$.

In the Euclidean field theory, we have the
generating functional of complete Schwinger functions. Actually,
the $(\lambda\varphi^{4})_{d}$ Euclidean theory is defined by
these Euclidean Green's functions. The Euclidean generating
functional $Z(h)$ is formally defined by the following functional
integral:
\begin{equation}
Z(h)=\int [d\varphi]\,\, \exp\left(-S+ \int d^{d}x\,
h(x)\varphi(x)\right),
\label{8}
\end{equation}
where $S=S_{0}+S_{I}$  is the action that usually describes a free scalar field with
the contributions $S_{0}$ and $S_{I}$ given respectively by
\begin{equation}
S_{0}(\varphi)=\int d^{d}x\, \left(\frac{1}{2}
(\partial\varphi)^{2}+\frac{1}{2}
m_{0}^{2}\,\varphi^{2}(x)\right),
\label{9}
\end{equation}
and
\begin{equation}
S_{I}(\varphi)= \int d^{d}x\,\frac{g_{0}}{4!} \,\varphi^{4}(x).
\label{10}
\end{equation}
The last integral is the interacting part, defined by the non-Gaussian contribution.
In Eq. (\ref{8}), $[d\varphi]$ is a translational invariant measure,
formally given by $[d\varphi]=\prod_{x} d\varphi(x)$. The terms
$g_{0}$ and $m_{0}^{2}$ are respectively the bare coupling constant
and the squared mass of the model. Finally, $h(x)$ is a smooth
function that we introduce to generate the Schwinger functions of
the theory by functional derivatives. As usual, the normalization factor is defined using the condition
$Z_{0}(h)|_{h=0}=1$. In the following, we are absorbing this normalization factor in the
functional measure.
In the weak-coupling perturbative expansion, we
perform a formal perturbative expansion with respect to the non-Gaussian
terms of the action. As a consequence of this formal expansion,
all the $n$-point unrenormalized Schwinger functions are
expressed in a powers series of the bare coupling
constant $g_{0}$ \cite{livron}.

It is well known that at non-zero temperatures the perturbative expansion in powers
of the coupling constant breaks down \cite{jakiw}. Therefore resummation schemes are necessary in order
to obtain reliable results. On of these resummation schemes is called the
CJT formalism. This formalism resums one-particle irreducible diagrams
to all orders. The stationary conditions for the effective action are the
Dyson-Schwinger equations, for the one and two-point correlation functions of the field theory model.
The equations for the two-point correlation functions in the Hartree-Fock approximation give us self-consistent
conditions or gap equations.

Let us define the generating functional for the Schwinger functions with the
usual source $h(x)$, but with an additional contribution $K(x,y)$ which couples
to $\frac{1}{2}(\varphi(x)\varphi(y))$.
\begin{equation}
Z(h,K)=\int [d\varphi]\,\, \exp\left(-S+ \int d^{d}x\,
h(x)\varphi(x)+\frac{1}{2}\int d^{d}x\int d^{d}y\, \varphi(x)K(x,y)\varphi(y)\right),
\label{cjt1}
\end{equation}
where $S$ is the action of the model. We can define the generating functional of connected Schwinger functions $W(h,K)$ defining
\begin{equation}
W(h,K)=\ln Z(h,K).
\label{cjt2}
\end{equation}
The normalized vacuum expectation value of the field $\varphi_{0}(x)$ and the connected two-point Schwinger
function $G_{c}(x,y)$ are given by
\begin{equation}
\frac{\delta W(h,K)}{\delta h(x)}=\varphi_{0}(x)
\label{cjt3}
\end{equation}
and
\begin{equation}
\frac{\delta W(h,K)}{\delta K(x,y)}=\frac{1}{2}\biggl(G_{c}(x,y)+\varphi_{0}(x)\varphi_{0}(y)\biggr).
\label{cjt4}
\end{equation}
Making use of a Legendre transformation, we obtain the effective action
\begin{equation}
\Gamma(\varphi_{0},G_{c})= W(h,K)-\varphi_{0}h-\frac{1}{2}\varphi_{0}K\varphi_{0}-\frac{1}{2}G_{c}K,
\label{cjt5}
\end{equation}
where we are using that
\begin{equation}
\varphi_{0}h=\int d^{d}x\, \varphi_{0}(x)h(x),
\label{cjt55}
\end{equation}
\begin{equation}
\varphi_{0}K\varphi_{0}=\int d^{d}x\,\int d^{d}y\, \varphi_{0}(x)K(x,y)\varphi_{0}(y).
\label{cjt555}
\end{equation}
and
\begin{equation}
\frac{1}{2}G_{c}K=\frac{1}{2}\int\,d^{d}x\,d^{d}y\,G_{c}(x,y)K(y,x).
\label{cjt6}
\end{equation}
Functional derivatives of the effective action $\Gamma(\varphi_{0},G_{c})$ give us
\begin{equation}
\frac{\delta\Gamma(\varphi_{0},G_{c})}{\delta\varphi_{0}(x)}=-h(x)-\int\,d^{d}y\,K(x,y)\varphi_{0}(y),
\label{cjt7}
\end{equation}
\begin{equation}
\frac{\delta\Gamma(\varphi_{0},G_{c})}{\delta\,G_{c}(x,y)}=-\frac{1}{2}K(x,y).
\label{cjt8}
\end{equation}
For vanishing sources, the stationary conditions which determines the
normalized expectation value of the field $\varphi_{0}(x)$ and the two-point Schwinger function are
\begin{equation}
\frac{\delta\Gamma(\varphi_{0},G_{c})}{\delta\varphi_{0}(x)}|_{_{\varphi_{0}=\varphi,G_{c}=G_{0}}}=0
\label{cjt7}
\end{equation}
and
\begin{equation}
\frac{\delta\Gamma(\varphi_{0},G_{c})}{\delta\,G_{c}(x,y)}|_{_{\varphi_{0}=\varphi,G_{c}=G_{0}}}=0.
\label{cjt9}
\end{equation}
The last equation corresponds to a Dyson-Schwinger equation for the full two-point Schwinger function.
The effective action $\Gamma(\varphi_{0}, G_{c})$ can be written as
\begin{equation}
\Gamma(\varphi_{0}, G_{c})=I(\varphi_{0})-\frac{1}{2}Tr(\ln\,G_{c}^{-1})-
\frac{1}{2}Tr(D^{-1}\,G_{c}-1)+\Gamma_{2}(\varphi_{0},G_{c}),
\label{cjt10}
\end{equation}
where $D^{-1}$ is the inverse of the tree-level two-point Schwinger function and
$\Gamma_{2}(\varphi_{0},G_{c})$ is the sum of all two-point irreducible diagrams where all lines
represent free propagators. Let us assume that $\varphi_{0}(x)=\varphi_{0}$, i.e., a constant field.
For such homogeneous system we can define the effective potential. Stationary condition give us a
Dyson-Schwinger equation.

\section{Replica field theory for disorder scalar field model}

\quad
Let us assume, for simplicity  that
we have only a quenched disordered field coupled with the scalar field. In this case,
one must obtain average values of extensive quantities. The question that now arises is to compute the
average value of the free energy.
Let us study the usual $d$-dimensional $\lambda\varphi^{4}$ model.
The Euclidean action associated with the model with disorder degrees of freedom can be written
as
\begin{equation}
S(\mu(x),\varphi(x))=\int d^{d}x\, \biggl(\frac{1}{2}
\varphi(x)\Bigl(-\Delta\,+m_{0}^{2}\Bigr)
\varphi(x)
+\frac{\lambda}{4!}\varphi^{4}(x)-\mu(x)\varphi(x)\biggr).
\label{dis1}
\end{equation}
The symbol $\Delta$ denotes the Laplacian in $\mathbb{R}^{d}$. There are different possibilities for the probability distributions.
Some of them are the following.
The random function $\mu(x)$ which is described by a Gaussian distribution,
has a general probability distribution of the form
\begin{equation}
P_{1}(\mu)=p_{1}\,exp\Biggl(-\frac{1}{2}\int\,d^{d}x\int\,d^{d}y
\,\mu(x)V^{-1}(x-y)\mu(y)\Biggr),
\label{dis22}
\end{equation}
which yields the following two-point correlation function
$\overline{\mu(x)\mu(y)} =  V(x-y)$. Another possibility is to
assume that the random variables characterizing the disorder exhibit no-long range correlations 
in the Euclidean $d$-dimensional manifold, therefore the
probability distribution is written as
\begin{equation}
P_{2}(\mu)=p_{2}\,exp\Biggl(-\frac{1}{2\,\sigma}\int\,d^{d}x(\mu(x))^{2}\Biggr).
\label{dis2}
\end{equation}
The quantity $\sigma$ is a small positive parameter associated with the disorder and $p_{0}$ is
a normalization constant.
In this case we have a delta correlated random field, with two-point correlation function $\overline{\mu(x)\mu(y)} =\sigma\delta^{d}(x-y)$.
We can also use the probability distribution $P_{3}(\sigma)$ defined as
\begin{equation}
P_{3}(\sigma)=p_{3}\delta\bigl(\mu(x)-\sigma\bigr)+(1-p_{3})\delta\bigl(\mu(x)+\sigma\bigr).
\label{ar}
\end{equation}
At this point, let us define the free energy $F(\mu)$ given by
\begin{equation}
F(\mu)=\ln\, \int[d\varphi]\exp\biggl(S(\varphi(x),\mu(x))\biggr).
\label{fe}
\end{equation}
In the presence of a quenched random field
$\mu(x)$, ground state configurations of the
field $\varphi(x)$ are defined by the saddle-point equation
\begin{equation}
-\Delta\varphi(x)+m_{0}^{2}\,\varphi(x)+\frac{\lambda}{3!}\varphi^{3}(x)=\mu(x).
\label{sa}
\end{equation}
The solutions of the above equation depends on particular configurations of the quenched fields.
As was discussed by many authors, perturbation theory is inappropriate to be used in systems where
the disorder defines a large number of local minima in the energy landscape. One way to
circumvented this problem is to integrate the random field using also the replica method.

The average value  of the free energy $F_{q}$ is given by
\begin{equation}
F_{q}=\int\,d\rho[\mu]\,F(\mu).
\label{sa27}
\end{equation}
In the expression that defines $F_{q}$ we are
averaging the free energy over random function $\mu(x)$ with the probability distribution $P_{2}(\mu)$.
We write $d\rho[\mu]=d[\mu] P(\mu)$, where $d[\mu]=\prod_{x} d\mu(x)$.  
Now we can give a precise definition of quenched variables. Quenched means that the probability distribution 
$d\rho[\mu]$ is not modified by the scalar field $\varphi(x)$. Therefore we have
\begin{equation}
F_{q}=\int\,d[\mu]P(\mu)\ln\int [d\varphi]\exp(-S(\varphi,\mu))
\label{sa2}
\end{equation}
The replica approach consists in the following steps. First we construct $Z^{\,n}=Z\times Z\times...\times Z$.
We interpret $Z^{\,n}$ as the partition function of a new system, formed from $n$ statistically
independent copies of the original system. Using the fact that $\ln Z= \lim_{n\rightarrow 0}\frac{Z^n-1}{n}$
we have that $\overline{\ln Z}=
\lim_{n\rightarrow 0}\frac{Z_{n}-1}{n}$, where $Z_{n}=\overline{Z^{\,n}}$.
The average value of the free energy in the presence of the quenched disorder
will be defined by
$F_{q}\equiv \lim_{n\rightarrow 0}\frac{Z_{n}-1}{n}$,
showing that the this quantity can be calculated from a zero-component
field theory, where a new system defined by $n$ replicas which are not more statistically independent.
Note that a procedure similar to an analytic continuation
must be used in the replica approach. First we define the replica partition
function for integer $n$. Next we extend this function to an analytic function of $n$ and
finally we take the limit when $n\rightarrow 0$.
We have  that
\begin{equation}
(Z(\mu))^{n}=\int\,\prod_{i=1}^{n}[d\varphi_{i}]\,\exp\biggl(-\sum_{i=1}^{n}S(\varphi_{i},\mu)\biggr)
\label{u1}
\end{equation}
Integration in the disorder, using the probability distribution $P_{2}(\mu)$, we obtain $Z_{n}$, where
\begin{equation}
Z_{n}=\int\,\prod_{i=1}^{n}[d\varphi_{i}]\,\exp\biggl(-S_{eff}(\varphi_{i},n)\biggr),
\label{aa11}
\end{equation}
and
\begin{equation}
S_{eff}(\varphi_{i},n)=\frac{1}{2}\sum_{i=1}^{n}\sum_{j=1}^{n}\int d^{d}x\int d^{d}y\,\,\varphi_{i}(x)
D_{ij}(x-y)
\varphi_{j}(y)+\frac{\lambda}{4!}\sum_{i=1}^{n}\int\,d^{d}x\, \varphi_{i}^{4}(x).
\label{aa12}
\end{equation}
In the above equation we have that $D_{ij}(x-y)=D_{ij}(m_{0},\sigma;x-y)$ where
\begin{equation}
D_{ij}(m_{0},\sigma;x-y)=\biggl(\delta_{ij}
(-\Delta+m_{0}^{2})-\sigma\biggr)\delta^{d}(x-y).
\label{aa124}
\end{equation}
Let us discuss the saddle-point equation of this model.
We have
\begin{equation}
\Bigl(-\Delta\,+m_{0}^{2}\Bigr)
\varphi_{i}(x)
+\frac{\lambda}{3!}\varphi^{3}_{i}(x)=\sigma\sum_{j=1}^{n}\varphi_{j}(x).
\label{sp}
\end{equation}
Let us suppose that $\varphi_{i}(x)=\varphi(x)$. In the limit $n\rightarrow\,0$ we obtain the pure system saddle-point which has the trivial
solution $\varphi(x)=0$ for $m^{2}>0$. Therefore for any non-trivial solution, the fields in different replicas cannot be equal.
The symmetry among replicas must be broken.

Let us discuss first the perturbative expansion of the replica field theory model.
Starting from a Gaussian theory, let us define $Z_{n}^{\,0}(h_{i})$
where we introduce an external source $h_{i}(x)$ which is a smooth
function that we introduce to generate the correlation  functions of
the theory by functional derivatives. We have

\begin{equation}
Z_{n}^{\,0}(h_{i})
=\int\prod_{i=1}^{n}[d\varphi_{i}]\exp{\biggl(-\frac{1}{2}
\sum_{i=1}^{n}\sum_{j=1}^{n}
\int\ d^{d}x\int d^{d}y
\,\varphi_{i}(x)
D_{ij}(x-y)\varphi_{j}(x)+\sum_{i=1}^{n}\varphi_{i}h_{i}
\biggr)},
\end{equation}
where we have that
\begin{equation}
\sum_{i=1}^{n}\varphi_{i}h_{i}=\sum_{i=1}^{n}\int d^{d}x\,h_{i}(x)\varphi_{i}(x).
\label{a111}
\end{equation}
Performing the gaussian integrals we get
\begin{equation}
Z_{n}^{\,0}(h_{i})
=\exp\biggl(\frac{1}{2}\sum_{i=1}^{n}\sum_{j=1}^{n}\int\,d^{d}x\int\,d^{d}y\,h_{i}(x)G_{ij}(m_{0},\sigma; x-y)h_{j}(y)\biggr),
\label{aa125}
\end{equation}
where $G_{ij}(m_{0},\sigma; x-y)$ is the two-point correlation function for the replica field theory. We have
\begin{equation}
\sum_{j=1}^{n}\int d^{d}z\,D_{ij}(x-z)
G_{jk}(m_{0},\sigma;z-y)=\delta_{ik}\,\delta^{d}(x-y).
\label{zz47}
\end{equation}
We can use $Z_{n}^{\,0}(h_{i})$ to obtain $Z_{n}$. We have
\begin{equation}
Z_{n}=\biggl[\exp\biggl(-\int\,d^{d}x\,V\biggl(\frac{\delta}{\delta\,h_{i}}\biggr)\Biggr)Z_{n}^{\,0}(h_{i})\biggr]|_{h_{i}=0}.
\label{aa126}
\end{equation}
We are assuming now a field theory model with replicas,
and that the finite-size effects effects must appear in the functional integral assuming some constrain in the functional space.
Instead of in coordinate space it is interesting
to give a treatment in momentum space. Performing a Fourier transform  we get
\begin{equation}
S_{eff}(\varphi_{i},n)=\frac{1}{2}\sum_{i=1}^{n}\sum_{j=1}^{n}\int\,\frac{d^{d}p}{(2\pi)^{d}}\,\,\varphi_{i}(p)\bigl[G_{0}
\bigr]_{ij}^{-1}\varphi_{j}(-p)+\frac{\lambda}{4!}\sum_{i=1}^{n}\varphi_{i}^{4},
\label{aa13}
\end{equation}
where in the quadratic part of $S_{eff}(\varphi_{i})$, the quantity $\bigl[G_{0}\bigr]_{ij}^{-1}$, which is the inverse of the
two-point correlation function in the tree-level approximation is defined as
\begin{equation}
\bigl[G_{0}\bigr]_{ij}^{-1}(p)=(p^{2}+m_{0}^{2})\delta_{ij}-\sigma.
\label{aa14}
\end{equation}
To invert this matrix, let us express $\bigl[G_{0}\bigr]_{ij}^{-1}$ in terms of
appropriate projector operators. We get
\begin{equation}
\bigl[G_{0}\bigr]_{ij}^{-1}(p)=(p^{2}+m_{0}^{2})\biggl(\delta_{ij}-\frac{1}{n}\biggr)+(p^{2}+m_{0}^{2}-n\sigma)\frac{1}{n}.
\label{aa15}
\end{equation}
The projectors operators, $(P_{T})_{ij}$ and $(P_{L})_{ij}$ are defined respectively as
\begin{equation}
(P_{T})_{ij}=\delta_{ij}-\frac{1}{n},
\label{aa16}
\end{equation}
and
\begin{equation}
(P_{L})_{ij}=\frac{1}{n}.
\label{aa17}
\end{equation}
Using the projectors operators we can write $\bigl[G_{0}\bigr]_{ij}(p)$ as
\begin{equation}
\bigl[G_{0}\bigr]_{ij}(p)=\frac{\delta_{ij}}{(p^{2}+m_{0}^{2})}+\frac{\sigma}{(p^{2}+m_{0}^{2})(p^{2}+m_{0}^{2}-n\sigma)}.
\label{aa18}
\end{equation}
As discussed by De Dominicis and Giardina \cite{livro4}, the first term is the bare contribution to the connected two-point correlation
function. The second term is the contribution to the disconnected two-point correlation 
function, which becomes connected after averaging the disorder.
In the next section we discuss finite-size effects in the one-loop approximation and also in the non-perturbative regime in the model with disorder.

\section{Finite-size effects in the one-loop approximation and non-perturbative results}

\quad

As it was discussed in the literature, one way to study the structure of the phase transition
of the random field Ising model is to adopt the
scalar field Landau-Ginzburg Hamiltonian density. The Hamiltonian of the system is defined as
\begin{equation}
{\cal{H}}_{LG}=\int\,d^{d}x\biggl(\frac{1}{2}(\nabla\varphi)^{2}+
\frac{1}{2}a\varphi^{2}+\frac{1}{4!}u\varphi^{4}-\mu(x)\varphi(x)\biggr),
\label{tem}
\end{equation}
where $a=a(T-T_{c})$. This quantity $a$ drives the phase transition in the model. We  will show that 
finite size effects acts as temperature effects in this classical system with disorder. Therefore it is possible to obtain a 
second order phase transition in one finite-size disordered $\lambda\varphi^{4}$ model, 
when the renormalized mass of the model will acquire finite size corrections.
The purpose of this section is to  present a detailed calculation of the one-loop renormalization of the
random field $\lambda\varphi^{4}$ model assuming that one space dimension is compactified,
and also obtain non-perturbative results. 

Before continue, let us discuss quantum field theory at finite temperature or with periodic boundary condition in one space dimension. 
Periodic boundary condition 
in one space dimension makes this field theory being defined in $S^{1}\times\,\mathbb{R}^{d-1}$ with the
Euclidean topology of a field theory at finite temperature. Nevertheless, for quantum systems with randomness, the situation that we are studying is  
not equivalent to finite temperature field theory. For such systems described by scalar fields, without randomness at thermal equilibrium with 
a reservoir and systems with one compactified dimension, finite size is totally equivalent to finite 
temperature field theory $(\beta=L)$ \cite{khanna}.
For systems with disorder degrees of freedom, this is not true anymore. In the case of finite temperature 
field theory,  with static disorder, the disorder degrees of freedom is uncorrelated 
in space but correlated in the Euclidean time direction. In this case it is necessary to use the two-point correlation function in the form 
\begin{equation}
\overline{\mu(x)\mu(x')} = \sigma V(\tau-\tau')\delta^{d-1}(x-x').
\label{nnn}
\end{equation}
This anisotropy in the behavior of the disorder in the Euclidean time direction show us that a quantum field theory with disorder degrees of freedom, finite temperature effects 
are not equivalent to finite size effects, discussed by us in this work. A different propose was presented in Refs. \cite{arias1,arias}.  
In both cases it was used the two-point correlation function in the form 
\begin{equation}
\overline{\mu(x)\mu(x')} = \sigma\delta^{d}(x-x').
\label{nnnn}
\end{equation}
to study the effects of light-cone fluctuations on the renormalized zero-point energy
associated with a free massless scalar field in the presence of boundaries and the thermodynamics of a relativistic charged scalar field in the
presence of disorder respectivelly.

Let us show that the correlation length is finite for small
radius of the compactified dimension. In order to find the
finite-size correction to the mass of the model we are using that all the Feynman rules are the same as in
the usual case, except that the momentum-space integrals over one component is
replaced by a sum over discrete frequencies. For the case of Bose fields with periodic boundary conditions we must perform the
replacement
\begin{equation}
\int\frac{d^{d}p}{(2\pi)^{d}}f(p)\rightarrow\,\frac{1}{L}\int\frac{d^{d-1}p}{(2\pi)^{d-1}}
\sum_{n=-{\infty}}^{\infty}f(\frac{2n\pi}{L},\vec{p}\,),
\label{aa19}
\end{equation}
where $L$ is the radius of the compactified dimension of the system and the vector $\vec{p}=(p^{2},p^{3},..,p^{d})$.
The two-point correlation function in the one-loop approximation can be written as
\begin{equation}
[G_{0}]_{lm}(x-y,L)=\mu^{4-d}\sum_{ij}\int\,d^{d}z\,[G_{0}]_{li}(x-z,L)[G_{0}]_{ij}(z-z,L)[G_{0}]_{jm}(z-y,L).
\label{sf}
\end{equation}
The functions $[G_{0}]_{li}(x-z,L)$ and $[G_{0}]_{jm}(z-y,L)$ are singular at $x=z$ and $y=z$, but the singularities
are integrable. The only contribution to the divergences is coming from $[G_{0}]_{ij}(z-z,L)$. Let us study the
quantities that appear in the definition of $[G_{0}]_{ij}(z-z,L)$.
At the one-loop approximation the unrenormalized size-dependent squared mass for the model is given by
\begin{equation}
m^{2}(L)=\,m_{0}^{2}+\delta m_{1}^{2}(L)+\delta m_{2}^{2}(L),
\label{aa20}
\end{equation}
where
\begin{equation}
\delta m_{1}^{2}(L)=\frac{\lambda}{2L}\int\frac{d^{d-1}p}{(2\pi)^{d-1}}
\sum_{n=-\infty}^{\infty}\frac{1}{\biggl((\frac{2\pi n}{L})^{2}+\vec{p}^{\,\,2}+m_{0}^{2}\biggr)},
\label{aa21}
\end{equation}
and
\begin{equation}
\delta m_{2}^{2}(L)=\frac{\lambda\sigma}{2L}\int\frac{d^{d-1}p}{(2\pi)^{d-1}}
\sum_{n=-\infty}^{\infty}\frac{1}{\biggl((\frac{2\pi n}{L})^{2}+\vec{p}^{\,\,2}+m_{0}^{2}\biggr)^{2}}.
\label{aa22}
\end{equation}
The integral appearing in  $\delta m_{1}^{2}(L)$ and $\delta m_{2}^{2}(L)$ can be calculated
using dimensional regularization \cite{analit1,analit2,analit3,analit4,analit5}. 
In the following we are using a mix between analytic and dimensional
regularization. It is well known the
formula:
\begin{equation}
\int d^{d}q\frac{1}{(q^{2}+a^{2})^{s}}=
\frac{\pi^{\frac{d}{2}}}{\Gamma(s)}
\Gamma\biggl(s-\frac{d}{2}\biggr)\frac{1}{(a^{2})^{s-\frac{d}{2}}}.
\label{dimreg}
\end{equation}
Using the dimensional regularization expression  we
obtain
\begin{equation}
\delta m_{1}^{2}(L)=\frac{\lambda}{2L}\frac{1}{(2\sqrt{\pi})^{d-1}}
\Gamma\biggl(\frac{3-d}{2}\biggr)\sum_{n=-\infty}^{\infty}\frac{1}{\biggl((\frac{2\pi n}{L})^{2}+m_{0}^{2}\biggr)^\frac{3-d}{2}}
\label{aa23}
\end{equation}
and
\begin{equation}
\delta m_{2}^{2}(L)=\frac{\lambda\sigma}{2L}\frac{1}{(2\sqrt{\pi})^{d-1}}
\Gamma\biggl(\frac{5-d}{2}\biggr)\sum_{n=-\infty}^{\infty}\frac{1}{\biggl((\frac{2\pi n}{L})^{2}+m_{0}^{2}\biggr)^\frac{5-d}{2}}.
\label{aa23}
\end{equation}
First, note that since we are using dimensional regularization there is implicit in the
definition of the coupling constant a factor $\mu^{4-d}$.
After use dimensional regularization we have to analytically extend the modified Epstein
zeta function. A well suited representation for the analytic extension is given in Ref. \cite{eli}. This zeta function is defined as
\begin{equation}
E(s,a)=\sum_{n=-\infty}^{\infty}\frac{1}{(n^{2}+a^{2})^{s}},
\label{epst}
\end{equation}
which converges absolutely and uniformly for $Re(s)>\frac{1}{2}$. A useful representation of the analytic extension
of this function is
 \begin{equation}
E(s,a)=\frac{\sqrt{\pi}}{\Gamma(s)a^{2s-1}}\biggl[\Gamma\biggl(s-\frac{1}{2}\biggr)
+4\sum_{n=1}^{\infty}(n\pi a)^{s-\frac{1}{2}}K_{s-\frac{1}{2}}(2\pi na)\biggr],
\label{epst2}
\end{equation}
where $K_{\nu}(z)$ is the modified Bessel function of third kind.
Now, the gamma function $\Gamma(z)$ is a
meromorphic function of the complex variable $z$ with simple poles
at the points $z=0,-1,-2,..$
In the neighborhood of any of
its poles $z=-n$, for $n=0,1,2,..$ $\Gamma(z)$ has a
representation given by
\begin{equation}
\Gamma(z)=\frac{(-1)^{n}}{n!}\frac{1}{(z+n)}+\Omega(z+n),
\label{gam}
\end{equation}
where $\Omega(z+n)$ stands for the regular part of the analytic
extension of $\Gamma(z)$. Both expression has a $L$-independent polar part plus a $L$-dependent analytic correction.
The mass counterterm, i.e., the principal part of the Laurent series of the analytic regularized quantity, is $L$-independent.
We are using a modified minimal subtraction renormalization scheme.
In conclusion, to deal with the ultraviolet divergences of the one-loop two-point correlation function we
use dimensional regularization to deal with the $(d-1)$ dimensional integrals and analytic regularization to deal with the
frequency sums. We get finally that the $L$-dependent renormalized squared mass is given by
\begin{equation}
m^{2}(L)=\,m_{0}^{2}+\lambda\,f_{1}(L,m_{0})-\lambda\sigma\,f_{2}(L,m_{0}),
\label{aa25}
\end{equation}
where $f_{1}(L,m_{0})$ and $f_{2}(L,m_{0})$ are given respectively by,
\begin{equation}
f_{1}(L,m_{0})=
\frac{1}{(2\pi)^{d/2}}\sum_{n=1}^{\infty}\biggl(\frac{m_{0}}{n L}\biggr)^{\frac{d}{2}-1}K_{\frac{d}{2}-1}(m_{0}nL),
\label{aa24}
\end{equation}
and
\begin{equation}
f_{2}(L,m_{0})=
\frac{3}{2(2\pi)^{d/2}}\sum_{n=1}^{\infty}\biggl(\frac{m_{0}}{n L}\biggr)^{\frac{d}{2}-2}K_{\frac{d}{2}-2}(m_{0}nL).
\label{aa24}
\end{equation}
It is clear that, for any $\sigma\neq 0$
there is a critical size $L_{c}$ where the system develop a second-order phase transition
where the system presents long-range correlations with
power-law decay. We obtain that at zero temperature, the $L$-dependent renormalized squared mass
can vanishes for some critical size $L_{c}$.
A  natural continuation is to see how this scheme works to two loops. There are other possibility.

\begin{figure}[htbp]
\centering
\subfigure[\tiny{Renormalized squared mass in d=3, as a function of the disorder parameter $\sigma$ and the radius of the compactified dimension $L$
          ($m_{0}=10$), in the one-loop approximation.}]{\includegraphics[scale=0.6]{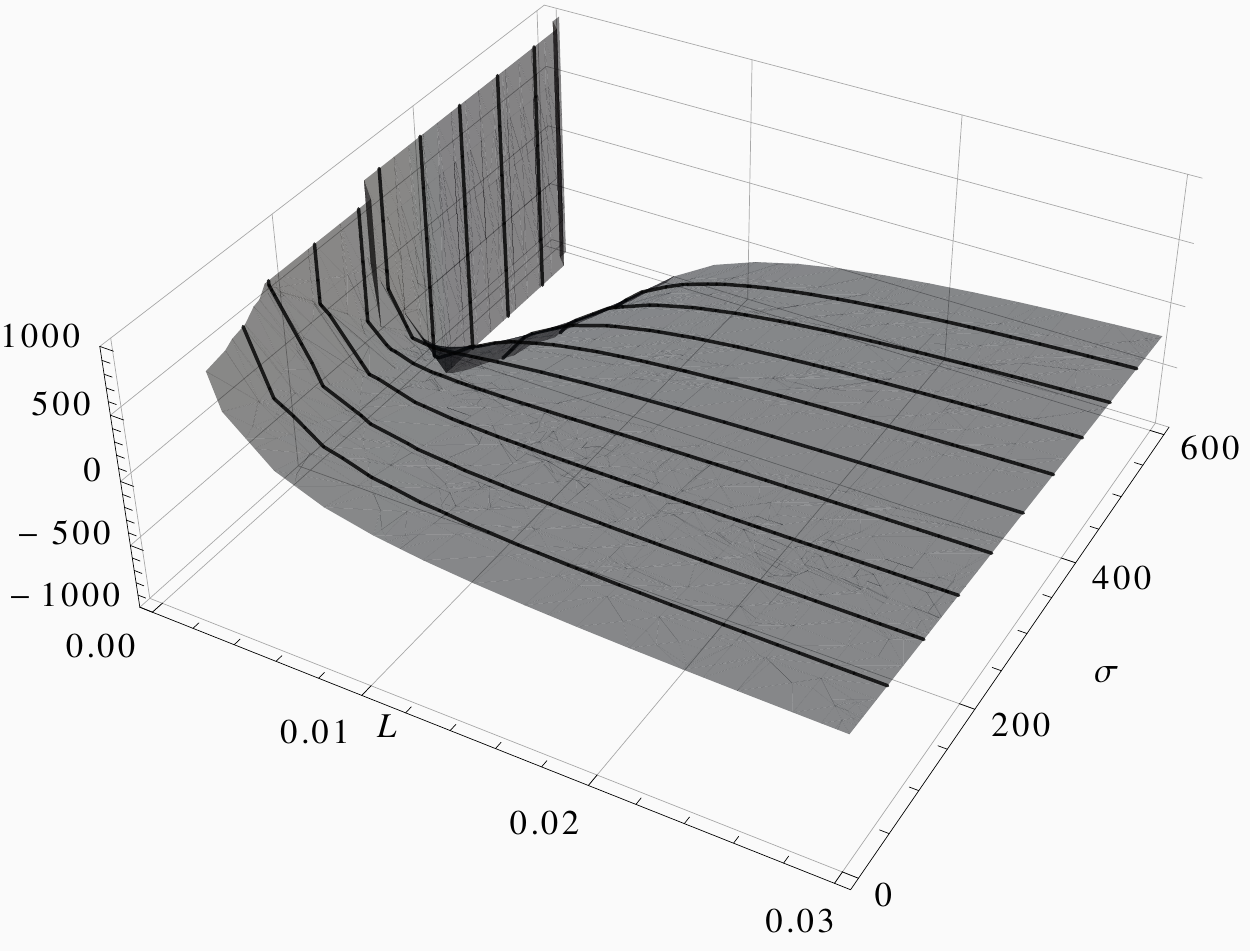}}\,\,\,\,\,\,\,\,\,\,\,\,\,
\subfigure[\tiny{Renormalized squared mass in d=4, as a function of the disorder parameter $\sigma$ and the radius of the compactified dimension $L$ 
          ($m_{0}=10$), in the one-loop approximation.}]{\includegraphics[scale=0.6]{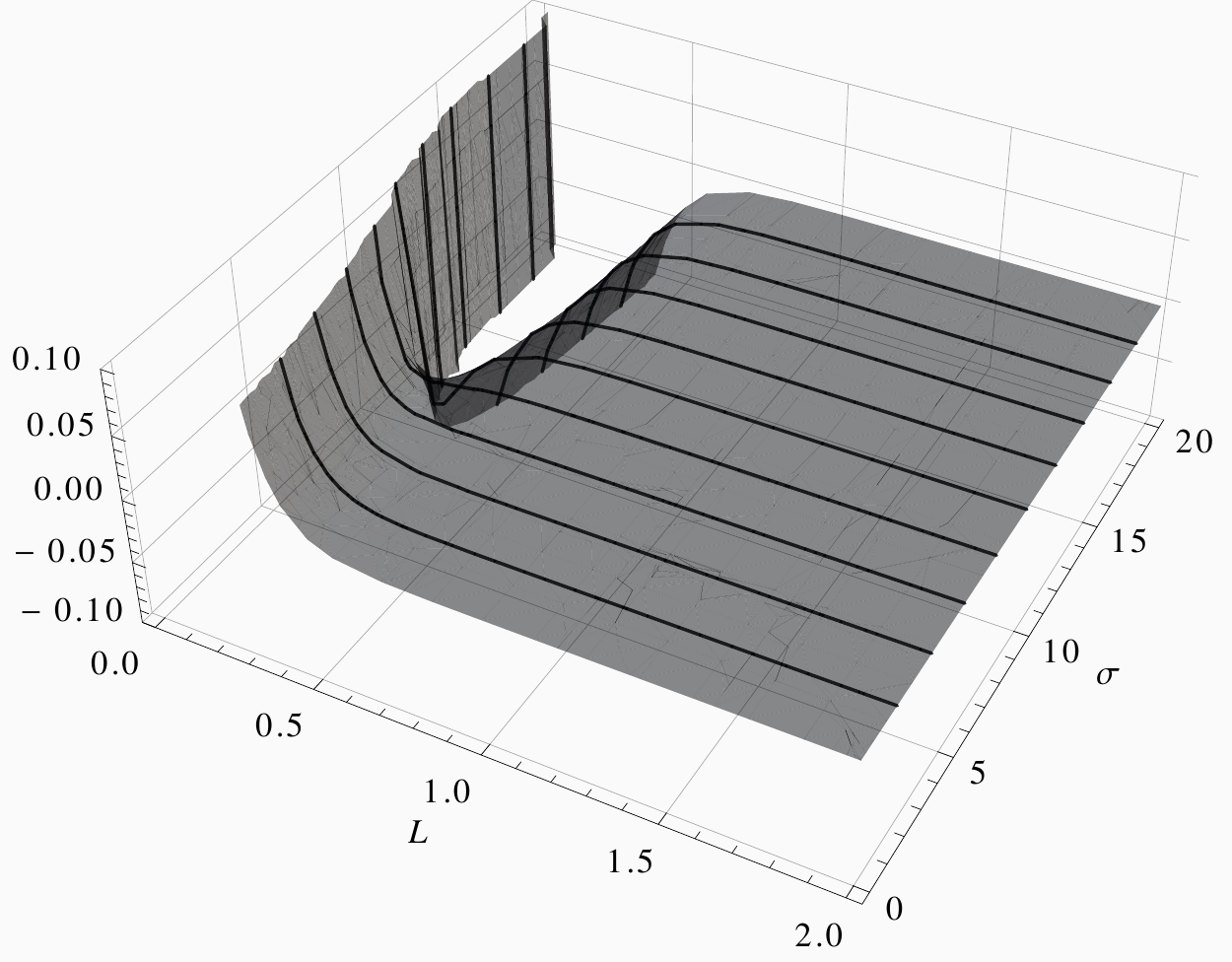}}
\end{figure}          
          
We can obtain nonperturbative results using the composite operator
formalism \cite{jackiw,ananos,gap} or the Dyson-Schwinger equations.
One way to implement the ressumation procedure is to write
\begin{equation}
m^{2}(\,L)=\delta m_{1}^{2}(L)+\delta m_{2}^{2}(L),
\label{cjt13}
\end{equation}
where
\begin{equation}
\delta m_{1}^{2}(L)=\frac{\lambda}{2L}\int\frac{d^{d-1}p}{(2\pi)^{d-1}}\sum_{n=-\infty}^{\infty}\frac{1}
{\biggl((\frac{2\pi n}{L})^{2}+\vec{p}^{\,\,2}+m
^{2}(L)\biggr)},
\label{cjt11}
\end{equation}
and
\begin{equation}
\delta m_{2}^{2}(L)=\frac{\lambda\sigma}{2L}\int\frac{d^{d-1}p}{(2\pi)^{d-1}}\sum_{n=
-\infty}^{\infty}\frac{1}{\biggl((\frac{2\pi n}{L})^{2}+\vec{p}^{\,\,2}+m^{2}(L)\biggr)^{2}}.
\label{cjt12}
\end{equation}
Using again, dimensional regularizarion and an analytic regularization procedure we obtain that
the gap equation that defines the size-dependent renormalized squared mass is given by
\begin{equation}
m^{2}(L)=\lambda\,f_{1}(L,m(L))-\lambda\sigma\,f_{2}(L,m(L)),
\label{aa25}
\end{equation}
where $f_{1}(L,m(L))$ and $f_{2}(L,m(L))$ are given respectively by,
\begin{equation}
f_{1}(L,m(L))=
\frac{1}{(2\pi)^{d/2}}\sum_{n=1}^{\infty}\biggl(\frac{m(L)}{n L}\biggr)^{\frac{d}{2}-1}K_{\frac{d}{2}-1}(m(L) nL),
\label{aa241}
\end{equation}
and
\begin{equation}
f_{2}(L,m(L))=
\frac{3}{2(2\pi)^{d/2}}\sum_{n=1}^{\infty}\biggl(\frac{m(L)}{n L}\biggr)^{\frac{d}{2}-2}K_{\frac{d}{2}-2}(m(L) nL).
\label{aa241}
\end{equation}
We obtained that for $\sigma\,>\,\sigma_{0}$ the renormalized squared mass is a monotonically decrescent function of the 
lenght $L$. When $m^{2}(L)=0$, this system presents long-range correlation with power-law decay.
Note that it is possible to investigate the behavior of the $L$-dependent correction to the coupling constant
of the model.
In the usual $\lambda\varphi^{4}$ model it was obtained that in $d=3$ the renormalized $L$-dependent coupling constant decreases up
to a minimum value different from zero and them grows monotonically with the inverse of the length. In the $d=4$ case
the renormalized $L$-dependent coupling constant is always a positive quantity. Therefore, the behavior
of the coupling constant with the radius of the compactified dimension is not able to change the structure of the phase transition of the model,
obtained by us in this work.

\begin{figure}[htbp]
\centering
\subfigure[\tiny{Renormalized squared mass in d=3, as a function of the disorder parameter $\sigma$ and the radius of the compactified dimension 
  $L$, using the composite operator method.}]{\includegraphics[scale=0.6]{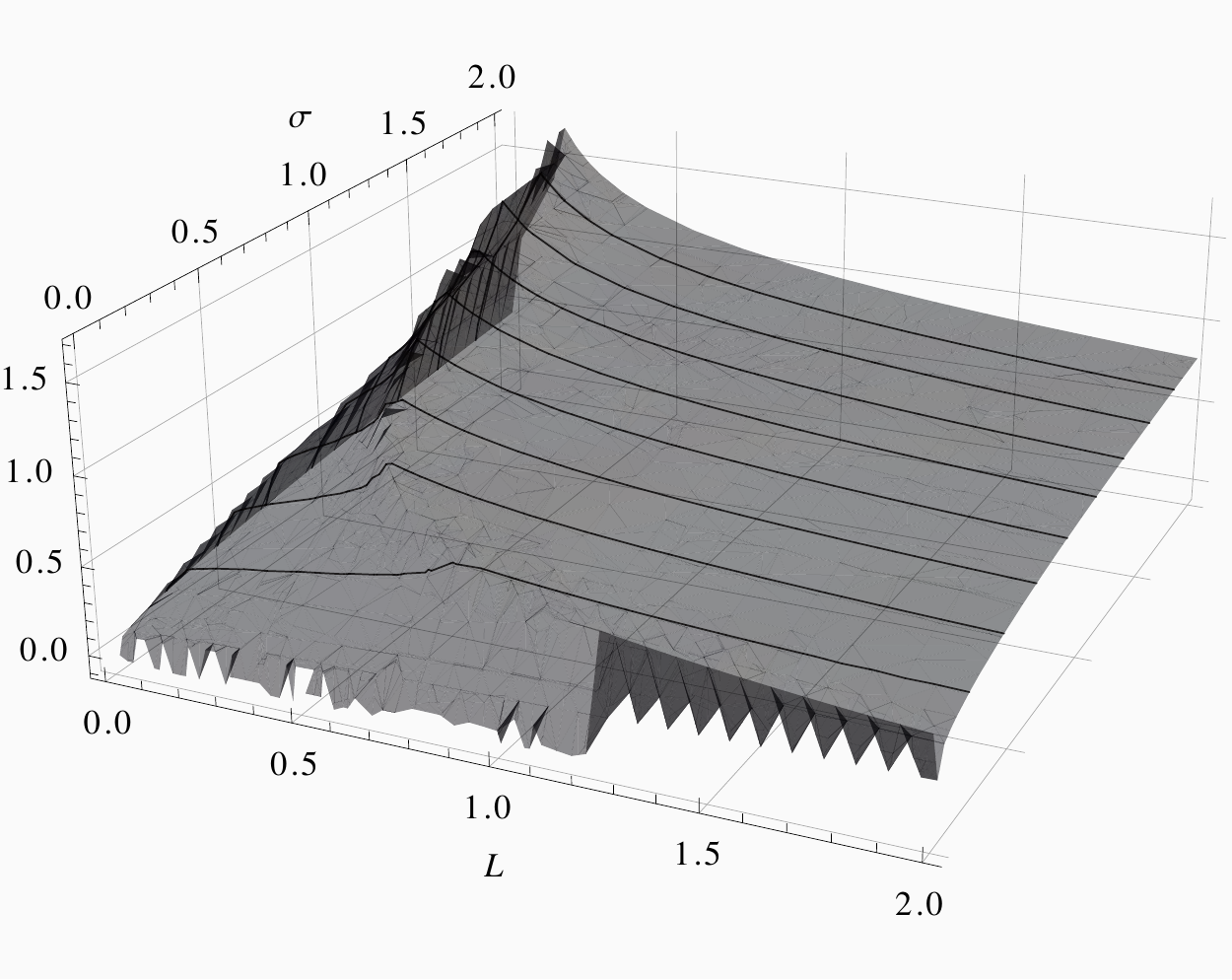}}\,\,\,\,\,\,\,\,\,\,\,\,\,
\subfigure[\tiny{Renormalized squared mass in d=4, as a function of the disorder parameter $\sigma$ and the radius of the compactified dimension  
  $L$, using the composite operator method.}]{\includegraphics[scale=0.6]{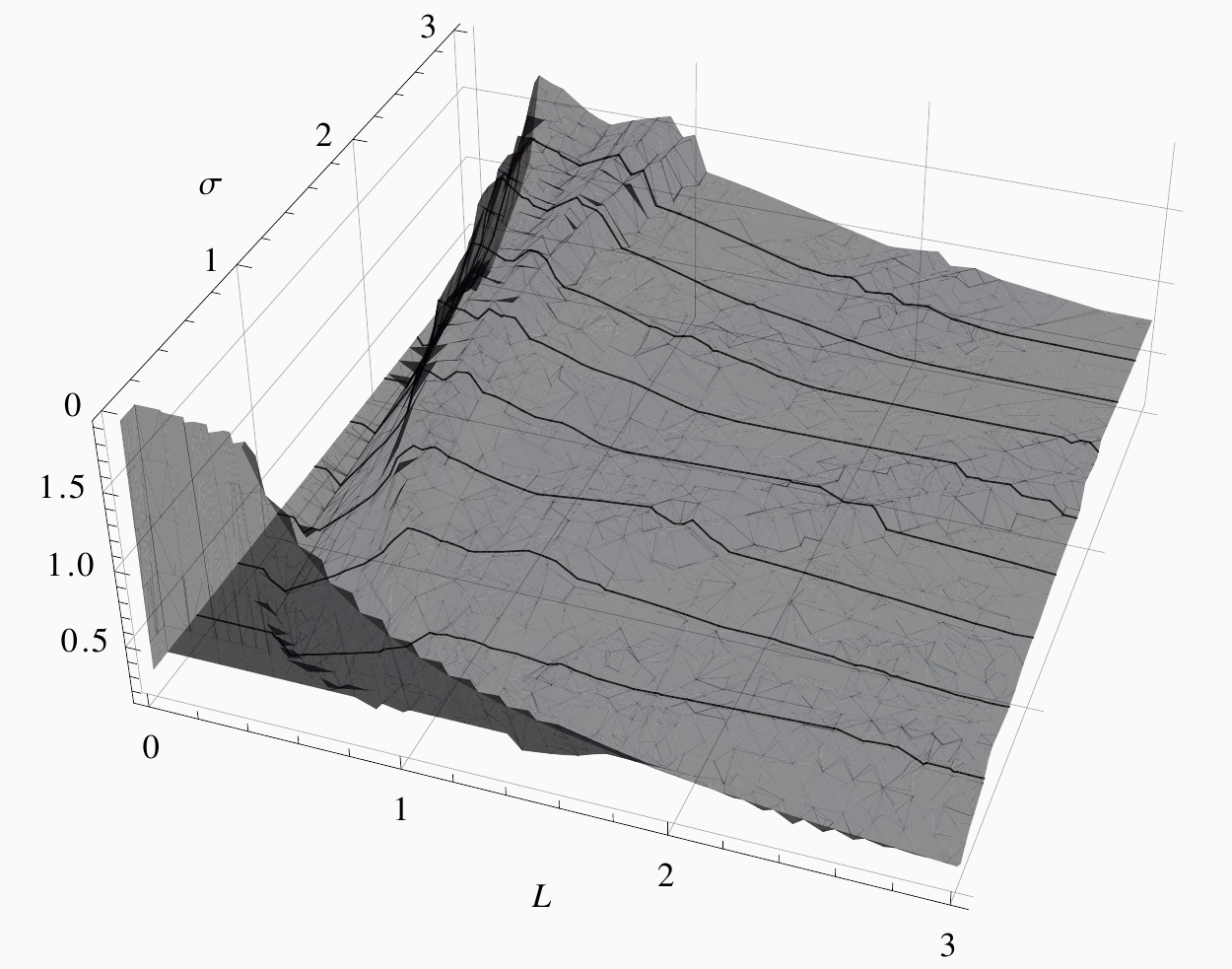}}
\end{figure}

\newpage

To conclude , we would like to discuss the infrared divergences. Direct application of field theoretical methods in the study of critical phenomena
has a long story. In the calculations of the critical exponents of different models one find non-integrable singularities in the integrals. In the 
critical region, it appears infrared divergences. One way to get around this problem is to enclose the system in a finite box and treat the zero 
mode properly. The behavior of the renormalized squared mass (size-dependent) in $d=3$ and $d=4$ is monotonically decrescent function of $L$.


\section{Conclusions}

\quad

In the last years, many efforts have been done to show that the universal problems in the theory of phase transitions 
can be formulated in the language of Euclidean field theory. The generating functional of Schwinger functions given by Eq. (\ref{8})
can be interpreted as the partition function of a $d$-dimensional 
classical statistical field theory. The zero temperature limit
of this theory corresponds to the usual infinite volume classical partition
function. Finite temperature effects in the generating functional of Schwinger functions
corresponds to a finite size $L = \beta$ in one direction for the classical partition
function. Our aim here is to use this Euclidean formalism of field theory 
to study classical statistical systems with disorder.

Suppose a scalar quantum field theory at zero temperature, without randomness. If we assume that the system obeys 
periodic boundary condition in one space dimension,
for sufficiently small radius, since there is a size-dependent renormalized squared 
mass, the system
must present finite correlation length.
This mass generation in the $\lambda\varphi^{4}$ theory has also been noted and studied by 
Ford and Yoshimura \cite{yoshimura} and Birrel and Ford \cite{birrel}.
This topological mass generation drives phase transitions in scalar field models.
Here in this work we show that this mechanism also works for systems with randomness. Using this topological generation of mass,
we discuss finite-size effects in one disordered $\lambda\varphi^{4}$ model in
a $d$-dimensional Euclidean space. The finite size-effects was introduced as boundary condition in the
path integral. We would like to stress that
this program to study finite size effects in models with randomness is not new in the literature.
For example, in the Sherington-Kirkpatrick spin glass, finite size effects in the model was investigated in
Refs. \cite{s1,s2}. Also, more recently finite size-effects  was discussed in two papers
\cite{ricci1,ricci2}. In these papers the authors study finite-size corrections to the free energy density in disordered
spin systems on sparse random graphs.

Here using the replica method, we study
finite-size effects in the one-loop approximation in the disordered model.
We show that there is a critical length
where the system develop a second-order phase transition, with long-range correlations with
power-law decay. Finally, using the composite field operator
method, non-perturbative results are obtained. Note that in the case of finite temperature quantum
field theory,  with static disorder, the disorder degrees of freedom is uncorrelated 
in space but correlated in the Euclidean time direction. 
This anisotropy in the behavior of the disorder in the Euclidean time direction and space indicate that for quantum systems with disorder degrees of freedom, finite temperature effects 
are not equivalent to finite size effects.

A  natural continuation of this paper is to study temperature effects in the
random $\lambda\varphi^{4}$ model in a generic $d$-dimensional space using the composite operator formalism. We conclude the paper calling the
attention for the reader that Brezin and Dominicis \cite{brezin}
claim that in the random field
Landau-Ginzburg model in a generic $d$-dimensional space,  new interaction terms must be considered beyond the
usual quartic
interaction, here we study only the scalar field with a
$(\lambda\varphi^{4})$ self-interaction. The introduction of new interactions and the study of the
phase structure at low and high temperatures, i.e.,
the phase diagram of the model is under investigation by the authors.

\section{Acknowlegements}

We would like to acknowledge T. Micklitz, M. Continentino, S. Alves Dias, F. Nobre and
G. Menezes for the fruitful discussions.
This paper was supported by Conselho Nacional de
Desenvolvimento Cientifico e Tecnol{\'o}gico do Brazil (CNPq)


\begin{thebibliography}{99}

\bibitem{pe2} L. H. Ford, Phys. Rev. {\bf D51}, 1692 (1995).
\bibitem{pe10} L. H. Ford and N. F. Svaiter, Phys. Rev. {\bf D56}, 2226 (1997).
\bibitem{pe11} L. H. Ford and N. F. Svaiter, Phys. Rev. {\bf D54}, 2640 (1996)
\bibitem{dis1} G. Krein, G. Menezes and N. F. Svaiter, Phys. Rev. Lett. {\bf 105}, 131301 (2010).
\bibitem{dis2} G. Krein, G. Menezes, E. Arias and N. F. Svaiter, Int. Jour. Mod. Phys. {\bf 27A}, 1250129 (2012).
\bibitem{dis3} C. G. H. Bessa, J. G. Duenas and and N. F. Svaiter, Class. Quant. Grav. {\bf 29}, 215011 (2012).
\bibitem{larkin} A. Larkin, Sov. Phys. JETP {\bf 31}, 784 (1970).
\bibitem{binder} K. Binder and A. P. Young, Rev. Mod. Phys. {\bf 58}, 801 (1986).
\bibitem{livro1} M. Mezard, G. Parisi and M. Virasoro, {\em{``Spin-Glass Theory and Beyond"}}, World Scientific, (1987).
\bibitem{livro2} T. Nattterman, in {\em{``Spin-Glasses and Random Fields"}}, A. P. Young (Editor), World Scientific (1988).
\bibitem{livro3} V. Dotsenko, {\em{``Introduction to the Replica Theory in Disordered Statistical Systems"}},
Cambridge University Press (2001).
\bibitem{livro4} C. De Dominicis and I. Giardina, {\em{``Random Fields and Spin Glass"}}, Cambridge University Press (2006).
\bibitem{afm} J. F. Fernandes, Europhys. Lett. {\bf 5}, 129 (1988).
\bibitem{fluids1} P. G. de Gennes, J. Phys. Chem {\bf 88}, 6469 (1984).
\bibitem{fluids2} T. MacFarland, G. T. Barkema and J. F. Marko, Phys. Rev. {\bf B53}, 148 (1996).
\bibitem{ma} Y. Imry and S. -K. Ma, Phys. Rev. Lett. {\bf 35}, 1399 (1975).
\bibitem{sourlas} G. Parisi and N. Sourlas, Phys. Rev. Lett. {\bf 43}, 744 (1979).
\bibitem{parisi1} G. Parisi, {\em{``An introduction to the statistical mechanics of amorphous systems"}}, in {\em{``Field Theory, Disorder and Simulations"}},
Word Scientific, Singapore (1992).
\bibitem{bric1} J. Bricmont and A. Kupiainen, Phys. Rev. Lett. {\bf 59}, 1829 (1987).
\bibitem{bric2} J. Bricmont and A. Kupiainen, Comm. Math. Phys. {\bf 116}, 539 (1988).
\bibitem{wehr1} M. Aizenman and T. Wehr, Phys. Rev. Lett. {\bf 62}, 2503, (1989).
\bibitem{mezard1} M. M\'ezard and A. P. Young, Europhys. Lett. {\bf 18}, 653 (1992).
\bibitem{mezard2} M. Mezard and R. Monasson, Phys. Rev. {\bf B50}, 7199 (1994).
\bibitem{dotsenko} V. Dotsenko, A. B. Harris, D. Sherington and R. B. Stinchcombe, J. Phys. {\bf A28}, 3093 (1995).
\bibitem{orland} C. De Dominicis, H. Orland and T. Tenesvari, J. Phys. I France {\bf 5}, 987 (1995).
\bibitem{dotsenko2} V. Dotsenko and M. M\'ezard, J. Phys. {\bf A30}, 3363 (1997).
\bibitem{dotsenko3} V. Dotsenko, J. Phys. {\bf A32}, 2949 (1999).
\bibitem{sherington1} F. Krzakala, F. Ricci-Tersengui and L. Zdeborav\'a, Phys. Rev. Lett. {\bf 104}, 207208 (2010).
\bibitem{sherington} F. Krzakala, F. Ricci-Tersengui, D. Sherington and L. Zdeborav\'a, J. Phys. {\bf A14}, 042003 (2011).
\bibitem{edwards} S. F. Edwards and P. W. Anderson, J. Phys. {\bf F5}, 965 (1975).
\bibitem{danchev} J. G. Brankov, D. M. Danchev and N. S. Tonchev {\em{``Theory of Critical Phenomena in Finite-Size Systems"}}, World Scientific, Singapure (2000).
\bibitem{nami} E. Arias, E. Goulart, G. Krein, G. Menezes and N. F. Svaiter,
Phys. Rev. {\bf D83}, 125022 (2011).
\bibitem{yoshimura} L. H. Ford and T. Yoshimura, Phys. Lett. {\bf 70A}, 89 (1979).
\bibitem{birrel} N. D. Birrel and L.H. Ford, Phys. Rev. {\bf D22}, 330 (1980).
\bibitem{s1} G. Parisi, F. Ritort and F. Slanina, J. Phys. {\bf A26}, 247 (1993).
\bibitem{s2} G. Parisi, F. Ritort and F. Slanina, J. Phys. {\bf A26}, 3775 (1993).
\bibitem{ricci1} C. Lucibello, F. Morone, G. Parisi, F. Ricci-Tersenghi and Tommaso Rizzo, Phys. Rev. {\bf E90}, 012146 (2014).
\bibitem{ricci2} U. Ferrari, C. Lucibello, F. Morone, G. Parisi, F. Ricci-Tersenghi and T. Rizzo, Phys. rev. {\bf B88}, 184201 (2013).
\bibitem{finitez1}C. D. Fosco and N. F. Svaiter, Jour. Math. Phys. {\bf 42}, 5185 (2001).
\bibitem{finitez2} M. I. Caicedo and N. F. Svaiter, Jour. Math. Phys. {\bf 45}, 179 (2004).
\bibitem{finitez3} N. F. Svaiter, Jour. Math. Phys. {\bf45}, 4524 (2004).
\bibitem{finitez4} M. J. Aparicio Alcalde, G. H. Flores and N. F. Svaiter, Jour. Math. Phys. {\bf 47}, 052303 (2006).
\bibitem{ford} L. H. Ford and N. F. Svaiter, Phys. Rev. {\bf D51}, 6981 (1995).
\bibitem{adolfo1} A. P. C. Malbouisson and N. F. Svaiter, Physica {\bf A233}, 573 (1996).
\bibitem{khanna} F. C. Khanna, A. P. C. Malbouisson and E. A. Santana, Phys. Rep. {\bf 539}, 135 (2014).
\bibitem{jackiw} J. M. Corwall, R. Jackiw and E. Tomboulis, Phys. rev. {\bf D15}, 2428 (1974).
\bibitem{ananos} G. N. J. A\~na\~nos, A. P. C. Malbouisson and N. F. Svaiter, Nucl. Phys. {\bf B547}, 221 (1999).
\bibitem{gap} N. F. Svaiter,  Physica {\bf A285}, 493 (2000).
\bibitem{jpa} A. P. C. Malbouisson and J. M. C. Malbouisson, J. Phys. {\bf A35}, 2263 (2002).
\bibitem{kap} J. Kapusta, D. B. Reiss and S. Rudaz, Nucl. Phys. {\bf B263}, 207 (1985).
\bibitem{dru} I. T. Drummond, R. R. Hogan, P. V. Landshoff and A. Rebhan, Nucl. Phys. {\bf B524}, 579 (1998).
\bibitem{dirk} J. T. Lenaghan  and D. H. Rischke, J. Phys. {\bf G26}, 431 (2000).
\bibitem{novo1} G. Tarjus and M. Tissier, Phys. Rev. Lett. {\bf 93}, 267008 (2004).
\bibitem{novo2} P. Le Doussal and K. J. Wiese, Phys. Rev. Lett. {\bf 96}, 197202 (2006).
\bibitem{novo3} A. A. Fedorenko, Phys. Rev. {\bf E86}, 021131 (2012).
\bibitem{livron} J. Zinn-Justin, {\em{``Quantum Field Theory and Critical
Phenomena"}} Oxford University Press, N.Y. (1996).
\bibitem{jakiw} L. Dolan and R. Jackiw, Phys. Rev. {\bf D9}, 3320 (1974).
\bibitem{arias1} E. Arias, C. G. H. Bessa, J. G. Duenas, G. Menezes, and N. F. Svaiter, 
Int. Jour. Mod. Phys. {\bf A29}, 1450024 (2014).
\bibitem{arias}  E. Arias, G. Krein, G. Menezes and N. F. Svaiter, J. Phys. {\bf A48}, 495002 (2015).
\bibitem{analit1} C. G. Bollini and J. J. Giambiagi, Nuovo Cim. {\bf B12}, 20 (1972).
\bibitem{analit2} J. F. Ashmore, Nuovo Cim. Lett. {\bf 4}, 289 (1972).
\bibitem{analit3} G. 't Hooft and M. Veltman, Nucl. Phys. {\bf B44}, 189 (1972).
\bibitem{analit4} G. 't Hooft and M. Veltman, Nucl. Phys. {\bf B50}, 318 (1972).
\bibitem{analit5} G. Leibrandt, Rev. Mod. Phys. {\bf 47}, 849 (1975).
\bibitem{eli} E. Elizalde and A. Romeo, J. Math. Phys. {\bf 30}, 1133 (1989).
\bibitem{brezin} E. Br\'ezin and C. D. Dominicis, Europhys. Lett. {\bf 44} 13 (1998).

\end{thebibliography}
\end{document}